\documentclass{article}
\usepackage{algorithm}     
\usepackage{algorithmic}   
\usepackage{multirow}
\usepackage{spconf,amsmath,graphicx,subfigure}

\usepackage{hyperref}
\hypersetup{
    colorlinks=true,            
    linkcolor=blue,             
    filecolor=magenta,          
    urlcolor=cyan,             
    pdftitle={Overleaf Example},
    pdfpagemode=FullScreen,
    }
\usepackage{colortbl}
\newcommand{\xxc}[2]{{\color{black} #1}}

\title{Establishing Stochastic Object Models from Noisy Data via Ambient Measurement-Integrated Diffusion }
\name{Xiaoning Lei$^1$, Jianwei Sun$^{2}$, Wenhao Cai$^3$, Xichen Xu$^2$\sthanks{Corresponding author}, Yanshu Wang$^2$, Hu Gao$^2$}
\address{1 Contemporary Amperex Technology Co.Limited, Ningde, 352100, China\\
2 Shanghai Jiao Tong University, Shanghai, 200240, China\\
3 East China Normal University, Shanghai, 200062, China\\
leixn01@outlook.com, happysunsir@gmail.com, 51275901066@stu.ecnu.edu.cn,\\ neptune\_2333@sjtu.edu.cn, isaac\_wang@sjtu.edu.cn,\\
gao\_h@sjtu.edu.cn }

\begin{document}
%
\maketitle
\begin{abstract}
Task-based assessments of image quality (IQ) are critical for evaluating medical imaging systems, which must account for randomness, including anatomical variability. Stochastic object models (SOMs) provide a statistical description of such variability. However, conventional mathematical SOMs fail to capture realistic anatomy, while most data-driven approaches typically require clean data, which are rarely available in clinical tasks. To address this challenge, we propose an unsupervised method called \textbf{A}mbient \textbf{M}easurement-\textbf{I}ntegrated \textbf{D}iffusion (AMID) with noise decoupling, which establishes clean SOMs directly from noisy measurements. AMID introduces a measurement-integrated strategy aligning measurement noise with the diffusion trajectory, and explicitly models coupling between measurement and diffusion noise across steps, and thus an ambient-consistency loss is thus designed based on it to learn clean SOMs.  Experiments on real CT and mammography datasets show that AMID outperforms existing methods in generation fidelity and yields more reliable task-based IQ evaluation, demonstrating its potential for unsupervised medical imaging analysis.
\end{abstract}
\vspace{-8pt}
\section{Introduction}
\label{sec:intro}
\vspace{-8pt}



\textbf{\xxc{Motivation.}} 
\xxc{The goal of a medical imaging system is to produce images that support accurate clinical decisions. Therefore, image quality (IQ) should be assessed in a task-based manner, which requires accounting for all sources of variability, including variability of imaged objects. Stochastic Object Models (SOMs) address this by providing statistical representations capable of generating ensembles of anatomically realistic phantoms. While traditional SOMs rely on mathematical formulations, recent research increasingly establishes them from real experimental data, which accurately captures the statistical properties of biological tissues.}

\noindent \textbf{Development and Limitations.} 
Existing approaches for constructing SOMs can be broadly divided into two main categories: 
(\romannumeral1) \textit{Mathematical methods}, such as binary texture~\cite{abbey2008ideal} and lumpy objects~\cite{rolland1992effect}, which offer interpretability but fail to capture realistic variations. 
(\romannumeral2) \textit{Data-driven methods}, which leverage real imaging data to learn more realistic SOMs. These methods can be further divided into \textit{Clean-supervised approaches}~\cite{cheng2021applications,hung2023med}, which establish accurate SOMs but require clean measurements that are rarely available in clinical practice. 
Notably, in modalities such as CT, measurement noise can be reasonably approximated as Gaussian~\cite{gravel2004method,sharma2025detail,fartiyal2025dual}, providing a statistical basis for constructing SOMs directly from noisy measurements. 
Therefore, GAN-based methods such as ProAmGAN~\cite{zhou2022learning} and AmbientCycleGAN~\cite{xu2025ambient} attempt to learn from noisy data and enhance interpretability, but often suffer from mode collapse and limited diversity. 
Diffusion-based methods, the leading paradigm for image synthesis, include Ambient Denoising Diffusion GAN~\cite{xu2024ambientcyclegan}, Ambient Diffusion~\cite{daras2023ambient}, DDRM~\cite{kawar2022denoising}, and DPS~\cite{chung2022diffusion}. 
While these methods extend diffusion to noisy or corrupted data, they are either restricted to specific noise types or rely on pre-trained models using clean data. 
\textbf{In summary, existing approaches either rely on clean supervision or lack robustness when trained from noisy data. 
Therefore, an unsupervised diffusion framework that establishes realistic SOMs directly from noisy medical measurements remains essential.}

\noindent \textbf{AMID.} \xxc{
To address it, we propose the Ambient Measurement-Integrated Diffusion with Noise Decoupling (AMID), which establishes clean SOMs directly from noisy measurements. Experiments on CT and mammography confirm its effectiveness. Our contributions are summarized as follows:
\vspace{-10pt}

\begin{itemize}
\item To our knowledge, this work firstly  presents a pure diffusion framework that establishes clean SOMs {directly from noisy measurements} without any clean supervisionm in the medical scenario.
\vspace{-8pt}

\item To address the challenge of using measurement information to build clean SOMs, a {measurement-integrated strategy} is proposed, which incorporates measurement noise into the forward process and preserves intrinsic structural cues. 
\vspace{-10pt}

\item To address the interference of measurement noise, the interaction between diffusion and measurement noise is remodeled, and an {ambient-consistency loss} is thus introduced. It enables AMID to capture the distribution of measurement noise and establish clean SOMs from noisy measurements.

\end{itemize}
}

\begin{figure*}[htb]
    \centering
    \includegraphics[width=0.92\textwidth]{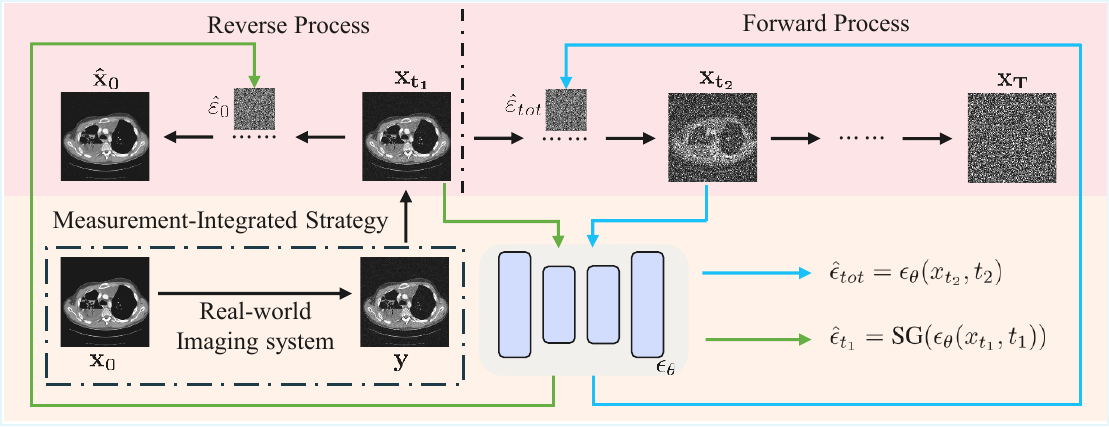}
    \vspace{-10pt}
    \caption{
        \normalsize
The architecture of the proposed AMID. The measurement $y$ is generated through the imaging system and thus contains noise. It is integrated into the diffusion process at a specific timestep $t_1$, after which it undergoes forward diffusion to step $T$. During inference, random Gaussian noise is first propagated to step $t_1$, 
and then the clean sample $x_0$ is directly obtained via DDIM sampling $p_\theta(x_0 \mid x_{t_1}) $
        }
        \vspace{-10pt}
    \label{fig:1}
\end{figure*}
\vspace{-20pt}
\section{Method}
\label{sec:method}
\vspace{-8pt}

\subsection{Preliminary studies}
\vspace{-5pt}
DDIM inference can be adopted to enable deterministic sampling between arbitrary timesteps; 
in particular, it allows directly mapping the intermediate state at $t_1$ to the clean sample $x_0$, as shown in Eq.~\ref{ddim_sam}. 
Thus, this mapping is used in AMID to estimate $x_0$ from the measurement-integrated intermediate state $x_{t_1}$.

\vspace{-25pt}
\begin{equation}
x_0 \approx \hat{x}_0(x_{t_1}) = \frac{1}{\sqrt{\bar\alpha_{t_1}}}\, x_{t_1} 
- \frac{\sqrt{1-\bar\alpha_{t_1}}}{\sqrt{\bar\alpha_{t_1}}}\,\hat\varepsilon_{t_1}.
\label{ddim_sam}
\end{equation}
\vspace{-15pt}

\vspace{-13pt}
\subsection{Measurement-Integrated Strategy}
\label{sec:sched}
\vspace{-5pt}

\noindent \xxc{Figure~\ref{fig:1} illustrates the framework of AMID. 
A noisy measurement is assigned to a specific step \(t_1\) of the diffusion process through the proposed measurement-integrated strategy, and then the diffusion proceeds to the final step \(T\). Moreover, the {ambient-consistency loss} models the joint effect of diffusion and measurement noise. It guides the model to capture the noise distribution and suppress residual artifacts. During inference, DDIM sampling is employed. It starts at step \(T\) and recovers a clean image at step \(t_1\), allowing clean SOMs to be obtained without requiring any clean supervision.}

\xxc{In a common description of a discrete imaging
system, the measurement $y$ can be modeled in Eq.\ref{eq1}:
\vspace{-6pt}
\begin{equation}
    y=\mathcal{H} x_0+\varepsilon_0,\quad \varepsilon_0\sim\mathcal{N}(0,\sigma_y^2I).
    \label{eq1}
    \vspace{-6pt}
\end{equation}
\noindent where $\varepsilon_0$ denotes the measurement noise. It can be reasonably approximated as Gaussian noise in most medical modalities. $\mathcal{H}$ is the imaging operator and $x_0$ is the underlying object. 

\noindent Moreover, in the forward diffusion process, $x_0$ is also gradually perturbed into pure noise by incrementally adding Gaussian noise, as shown in Eq.~\ref{eq2}.
\vspace{-6pt}

\begin{equation}
    x_t = {{\zeta}_t}\,x_0 + {\beta_t}\,\varepsilon_t,
    \quad \varepsilon_t \sim \mathcal{N}(0,I)
    \label{eq2}.
\end{equation}

\noindent where $\zeta_t = \sqrt{\bar\alpha_t}$ and $\beta_t = \sqrt{1-\bar\alpha_t}$. Following DMID~\cite{li2024stimulating}, when the measurement noise is approximated by a Gaussian distribution, $y$ can be aligned with the diffusion trajectory. The alignment enables the subsequent forward and reverse processes to inherently combine the imaging system with the diffusion chain, leading to the establishment of clean SOMs. 
\begin{algorithm}[t]
\caption{Measurement-Integrated Strategy}
\label{algo}
\textbf{Input:} Noisy measurement $y$, noise schedule $\{\bar{\alpha}_t\}_{t=1}^T$ \\
\textbf{Goal:} Intermediate step $t_1$ \\
\textbf{Procedure:} 
\begin{algorithmic}
\STATE Estimate noise standard deviation $\sigma_y \leftarrow \mathrm{std}(y)$
\STATE Normalize the measurement: 
$x'_{t_1} = \tfrac{y}{\sqrt{1+\sigma_y^2}} 
= \tfrac{1}{\sqrt{1+\sigma_y^2}} x_0 
+ \tfrac{\sigma_y}{\sqrt{1+\sigma_y^2}} \varepsilon,\;\varepsilon\sim\mathcal{N}(0,I)$
\STATE Find $t_1 = \arg\min_{t\in\{1,\dots,T\}} \big|\sqrt{\bar{\alpha}_t}-\tfrac{1}{\sqrt{1+\sigma_y^2}}\big|$
\STATE Approximate $x'_{t_1}$ as $x_{t_1}$ at step $t_1$
\STATE Forward diffusion: $x_t \to x_T, \;\; t=t_1,\dots,T$
\end{algorithmic}
\end{algorithm}

However, since the diffusion process assigns different coefficients $\zeta_t$ and $\beta_t$ to the image and noise components at each timestep $t$ (satisfying $\zeta_t^2 + \beta_t^2 = 1$), it is necessary to align the noise level of $y$ with that of the diffusion trajectory. Therefore, we propose a measurement-integrated strategy based on noise variance, as shown in Algorithm~\ref{algo}. Specifically, we estimate the noise variance of $y$ and normalize it to match the form of the forward diffusion process. The corresponding step $t_1$ is then determined by matching the normalized coefficient with the predefined $\{\sqrt{\bar{\alpha}_t}\}_{t=1}^{T}$. Once it is integrated at $t_1$, the sample proceeds through the standard forward diffusion to $T$, resulting in a latent state aligned with the trajectory. This strategy seamlessly aligns the noisy measurement $y$ with the forward trajectory and thereby facilitates denoising and the subsequent construction of SOMs without clean data.
}

\vspace{-28pt}
\subsection{Measurement–Diffusion Noise Decoupling}
\label{dec}
\vspace{-10pt}

\xxc{
Given a noisy measurement aligned at timestep $t_1$ in Sec.~\ref{sec:sched}, it is necessary to propagate it further to an arbitrary step $t_2>t_1$. Therefore, we define a propagation-consistent transition that analytically transports the latent from $t_1$ to $t_2$ within the diffusion forward chain, as shown in Eq.~\ref{eq3}.
\vspace{-16pt}
\begin{equation}
    x_{t_2} \;=\; \rho\,x_{t_1} \;+\; \sqrt{1-\rho^2}\,\varepsilon_{t_2},\qquad 
    \varepsilon_{t_2}\sim\mathcal{N}(0,I).
    \label{eq3}
\vspace{-18pt}
\end{equation}

\noindent where $\rho=\sqrt{\tfrac{\bar\alpha_{t_2}}{\bar\alpha_{t_1}}}$. By recursively expanding the forward diffusion transitions and grouping the Gaussian perturbations, $\varepsilon_{t_2}$ represents the stochastic diffusion noise injected between $t_1$ and $t_2$. $x_{t_2}$ is expressed as a deterministic scaling of $x_{t_1}$ plus an additional known noise $\varepsilon_{t_2}$. Considering the format of Eq.~\ref{eq1}, it is rewritten as:
\vspace{-10pt}
\begin{equation}
\begin{aligned}
x_{t_2} &= \sqrt{\bar\alpha_{t_2}}\,x_0 
        + \sqrt{1-\bar\alpha_{t_2}}\,\varepsilon_{\text{tot}} \\[4pt]
        &= \sqrt{\bar\alpha_{t_2}}\,x_0 
        + \sqrt{1-\bar\alpha_{t_2}}\,
          \left(\omega_1 \varepsilon_0 + \omega_2 \varepsilon_{t_2}\right)
\end{aligned}
\label{eq4}
\vspace{-15pt}
\end{equation}

\noindent where $\omega_1 = \frac{\rho\sqrt{1-\bar\alpha_{t_1}}}{\sqrt{1-\bar\alpha_{t_2}}}$ and $\omega_2 = \frac{\sqrt{1-\rho^2}}{\sqrt{1-\bar\alpha_{t_2}}}$. 
Here, $\varepsilon_0$ denotes the scaled measurement noise, which follows a standard Gaussian distribution
The total noise $\varepsilon_{\text{tot}}$ in the forward process remains Gaussian, since $\omega_1^2+\omega_2^2=1$.

\noindent Eq.~\ref{eq4} explicitly illustrates how intrinsic measurement noise and diffusion noise are combined with weights $(\omega_1,\omega_2)$, preserving measurement information along the diffusion trajectory. Thus, at convergence, $\varepsilon_{\text{tot}}$ and $\varepsilon_0$ correspond to network outputs at timesteps $t_2$ and $t_1$, respectively, allowing direct estimation of measurement noise for constructing clean SOMs and motivating the loss design in Sec.~\ref{ambient}.
}

\subsection{Ambient-consistency Loss}
\label{ambient}
Our training objective consists of two complementary terms:
\begin{equation}
\begin{aligned}
L_{1} &= 
||
  \hat\varepsilon_{\text{tot}}
  - \frac{\rho\sqrt{1-\bar\alpha_{t_1}}}{\sqrt{1-\bar\alpha_{t_2}}}
    \,\textit{SG}(\hat\varepsilon_{t_1})
  - \frac{\sqrt{1-\rho^2}}{\sqrt{1-\bar\alpha_{t_2}}}\,\varepsilon_{t_2}
||_2^2, \\[6pt]
L_{2} &= 
||
  \big(\hat\varepsilon_{\text{tot}}^{(a)} - \hat\varepsilon_{\text{tot}}^{(b)}\big)
  - \frac{\sqrt{1-\rho^2}}{\sqrt{1-\bar\alpha_{t_2}}}
    \big(\varepsilon_{t_2}^{(a)} - \varepsilon_{t_2}^{(b)}\big)
||_2^2, \\[6pt]
L &= L_{1} + \lambda L_{2},
\end{aligned}
\label{eq:loss}
\end{equation}
\vspace{-15pt}

\noindent where $\lambda$ balances the contributions of $L_2$. $L_1$ minimizes the gap between the predicted noise and the mixed target at step $t_2$ ($t_2 > t_1$). Here, $\textit{SG}(\cdot)$ denotes stop-gradient, which fixes $\hat\varepsilon_{t_1}$ as a reference for $\varepsilon_{0}$  and prevents unstable feedback. While $L_1$ alone suffices in theory, residual errors in $\hat\varepsilon_{t_1}$ propagate to later steps and appear as high-frequency noise. To mitigate this, $L_2$ introduces a difference loss by sampling two independent noises $\varepsilon^{(a)},\varepsilon^{(b)}$ at the same timestep. Since the shared component linked to $x_{t_1}$ cancels out, $L_2$ enforces consistency only along the noise direction, effectively reducing residual accumulation and suppressing artifacts.

\vspace{-10pt}
\section{EXPERIMENTS}
\label{sec:exp}
\vspace{-10pt}

\subsection{Datasets and Implementation Details}
\label{sec:datasets}

\vspace{-6pt}
We conduct experiments on two clinical datasets. The CT dataset, from DeepLesion~\cite{yan2018deeplesion}, includes over 22,000 images resized to 256$\times$256 with simulated measurements obtained by adding Gaussian noise (mean 0, std 0.06). The mammography dataset, from DDSM/CBIS-DDSM~\cite{heath1998current,lee2017curated}, contains 13,190 images resized to 256$\times$256, with Gaussian noise (mean 0, std 0.08) added for realism. For task-based image quality evaluation, we perform a signal-known-exactly (SKE) binary detection task, where 64×64 background patches from both ground truth and model outputs are equipped with Gaussian signals (width 0.3, amplitude 0.32) to simulate signal-present cases.
\vspace{-12pt}

\subsection{Comparison Studies}
\label{res}
\vspace{-6pt}
In our experiments, we compare our method with representative approaches trained on noisy data, including AmbientStyleGAN3~\cite{ambstylegan3}, DDPM~\cite{ho2020denoising}, the diffusion–GAN hybrid ADDGAN~\cite{xu2025ambient}, and the diffusion-based AmbientDiffusion~\cite{daras2023ambient}. Collectively, these baselines encompass state-of-the-art paradigms for establishing SOMs.

\begin{table}[h]
    \centering
    \resizebox{0.95\linewidth}{!}{
    \begin{tabular}{ccccc}
        \hline
        \multirow{2}{*}{Different methods} & \multicolumn{2}{c}{CT} & \multicolumn{2}{c}{DDSM} \\ \cline{2-5}
                                 & FID$\downarrow$ & IS$\uparrow$ & FID$\downarrow$ & IS$\uparrow$ \\ \hline
        DDPM               & 77.21  & 2.51 & 100.39 & 2.06 \\
        ADDGAN             & 53.97  & 2.67 & 64.45  & 2.28 \\
        AmbientDiffusion  & 144.87 & 2.57 & 151.61 & 1.71 \\ 
        AmbientStyleGAN3  & 168.85 & 2.36 & 305.76 & 2.09 \\ \hline
        AMID    & \textbf{40.38}  & \textbf{2.88} & \textbf{59.34}  & \textbf{2.31} \\ \hline
    \end{tabular}
    }
    \vspace{-5pt}
\caption{FID and IS on CT and DDSM datasets (8{,}000 generated and real images in each case).}
\label{tab:fid_is}
\end{table}

\vspace{-8pt}
\begin{figure}[htb]
\vspace{-10pt}
    \centering
    \includegraphics[width=0.47\textwidth]{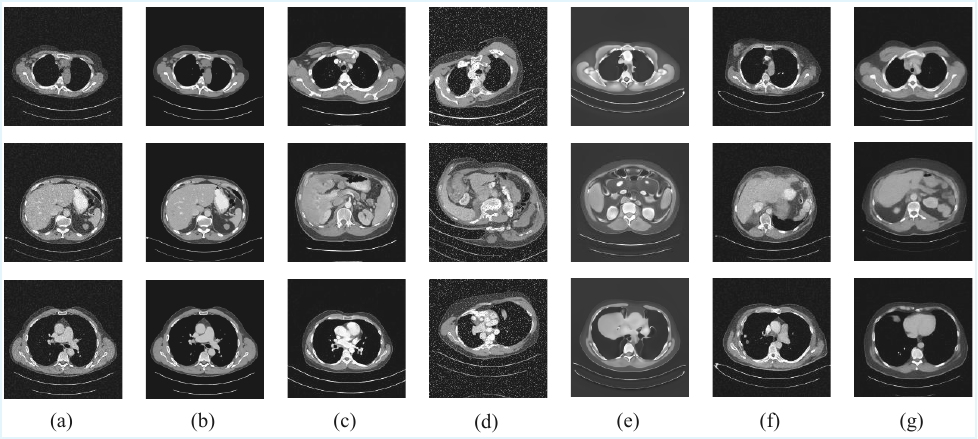}
    \vspace{-13pt}
   \caption{Qualitative comparison on the CT data. Columns correspond to (a) noisy input, (b) ground truth, (c) ADDGAN, (d) AmbientDiffusion, (e) AmbientStyleGAN3, (f) DDPM, and (g) the proposed AMID. }
    \label{fig:qualitative_CT}
\end{figure}
\vspace{-10pt}
\begin{figure}[htb]
    \centering
    \includegraphics[width=0.47\textwidth]{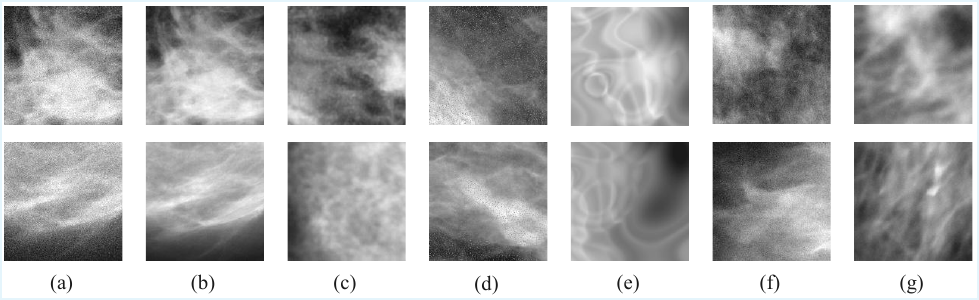}
    \vspace{-12pt}
    \caption{Qualitative comparison on the  mammography data. 
Columns correspond to (a) noisy input, (b) ground truth, (c) ADDGAN, (d) AmbientDiffusion, 
(e) AmbientStyleGAN3, (f) DDPM, and (g) the proposed AMID. }
    \label{fig:qualitative_DDSM}
    \vspace{-18pt}
\end{figure}

\begin{figure}[htb]
\vspace{-10pt}
    \centering
    \includegraphics[width=0.49\textwidth]{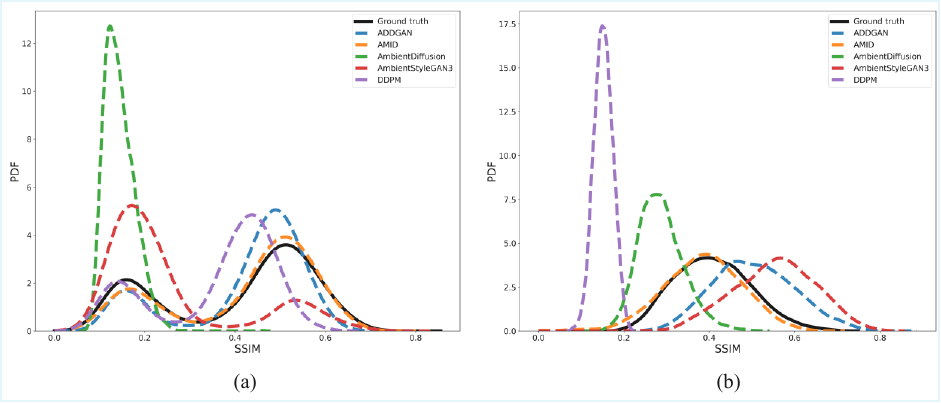}
    \vspace{-25pt}
    \caption{Comparison of SSIM-PDF between generated and ground-truth data. (a) Computed from 5,000 CT images. 
(b) Computed from 5,000 mammography images.}
    \label{ssim-ct}
\end{figure}

\noindent \textbf{Common IQ Evaluation.} The performance of the generative models is assessed using three metrics: the Fréchet Inception Distance (FID), the Inception Score (IS), and SSIM-PDF. SSIM-PDF is defined as the probability density function of the SSIM values computed between random pairs of ground-truth and model-generated patches. All results are summarized in Table~\ref{tab:fid_is}. The proposed method achieves the best scores on both datasets. On the CT dataset, it attains an FID of 40.38 and an IS of 2.88, outperforming the second-best method (ADDGAN) by 13.59 in FID and 0.21 in IS. On the DDSM dataset, it also performs best. As shown in Fig.~\ref{ssim-ct}, the SSIM-PDF curves of AMID are consistently closer to those of the ground truth, indicating that the generated images better preserve structural similarity. These results consistently confirm the advantages of the proposed method in producing images with higher fidelity and robust structural consistency. Moreover, as illustrated in Fig.~\ref{fig:qualitative_CT} and Fig.~\ref{fig:qualitative_DDSM}, AMID delivers the most visually convincing results: AmbientDiffusion and DDPM struggle with complex Gaussian noise, AmbientStyleGAN3 produces images that lack fine details, and ADDGAN introduces prominent artifacts in mammography images.

\vspace{-10pt}
\begin{figure}[htb]
    \centering
    \includegraphics[width=0.49\textwidth]{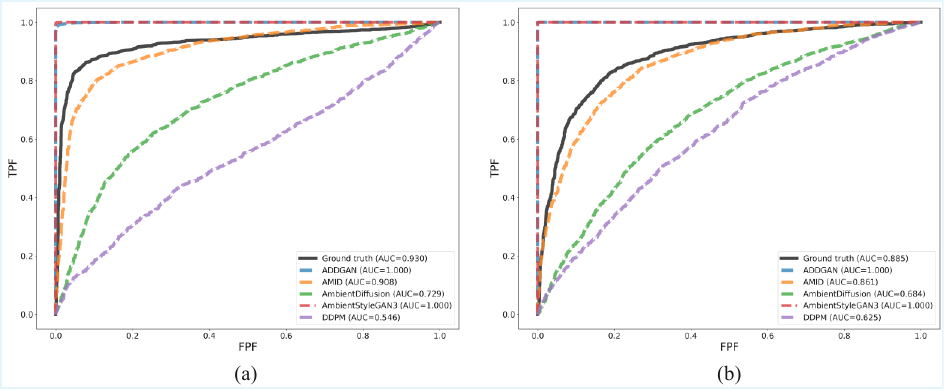}
    \vspace{-20pt}
\caption{Task-based IQ evaluation on 12,000 mammography patches using a signal-known-exactly (SKE) binary detection task. 
(a) Results with the Hotelling observer; 
(b) results with a ResNet50 model observer.}
    \label{ho-ddsm}
        \vspace{-16pt}
\end{figure}

\noindent \textbf{Task-based IQ Evaluation.} 
Beyond conventional metrics, we further conduct task-based IQ assessment to evaluate the suitability of generated images for downstream signal detection. Specifically, a signal-known-exactly (SKE) binary detection task is designed, where $64 \times 64$ background patches are trimmed from real and synthetic data, and Gaussian signals are added to form signal-present cases. Performance is quantified using two complementary observers: (\romannumeral1) the Hotelling observer, which maximizes the signal-to-noise ratio between the signal-present and signal-absent cases, and (\romannumeral2) a CNN-based model observer implemented with ResNet50, fine-tuned for the same SKE task. In total, 6{,}000 signal-present and 6{,}000 signal-absent patches are used for training, with 2{,}000 independent test patches for evaluation. As shown in Fig.~\ref{ho-ddsm}, AMID consistently outperforms baselines under both observers, demonstrating that the generated images preserve clinically relevant features and support reliable task-driven analysis.
\vspace{-8pt}

\begin{figure}[htb]
    \centering
    \includegraphics[width=0.47\textwidth]{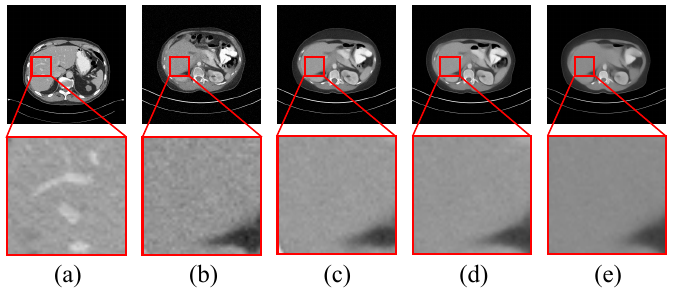}
    \vspace{-12pt}
    \caption{Effect of the trade-off parameter $\lambda$ on synthesized SOMs. A larger $\lambda$ suppresses high-frequency noise but oversmooths details, while $\lambda=0.2$ provides the best balance.}
    \label{abla}
\end{figure}

\vspace{-25pt}
\subsection{Ablation study}
\vspace{-8pt}

We further evaluate the impact of the trade-off parameter $\lambda$ in ambient-consistency loss, with values ${0,0.2,0.5,0.75}$. As shown in Fig.~\ref{abla}, increasing $\lambda$ effectively suppresses high-frequency noise in synthesized SOMs, resulting in smoother images. However, since $L_2$ only enforces consistency along the newly injected noise direction, larger $\lambda$ reduces the contribution of $\hat\varepsilon_{t_1}$, causing the structural cues from $x_{t_1}$ to be under-utilized and leading to a loss of fine details. In practice, setting $\lambda=0.2$ achieves a favorable balance, mitigating high-frequency noise while preserving both global structure and local details.

\vspace{-10pt}
\section{CONCLUSIONS}
\label{sec:conclude}
\vspace{-8pt}

In this study, we propose a novel \textbf{A}mbient \textbf{M}easurement-\textbf{I}ntegrated \textbf{D}iffusion (AMID) for learning high-quality stochastic object models (SOMs) directly from noisy measurements. By integrating measurements into the diffusion process and introducing the ambient-consistency loss, AMID generates realistic images without relying on clean training data or clean pre-trained models. Experiments demonstrate that AMID effectively establishes SOMs with statistical properties that closely match real tissues, offers a powerful tool for the evaluation of medical image quality.

\bibliographystyle{IEEEbib}
\clearpage
\bibliography{refs}

@inproceedings{abbey2008ideal,
  title={An ideal observer for a model of x-ray imaging in breast parenchymal tissue},
  author={Abbey, Craig K and Boone, John M},
  booktitle={International Workshop on Digital Mammography},
  pages={393--400},
  year={2008},
  organization={Springer}
}

@article{rolland1992effect,
  title={Effect of random background inhomogeneity on observer detection performance},
  author={Rolland, Jannick P and Barrett, Harrison H},
  journal={Journal of the Optical Society of America A},
  volume={9},
  number={5},
  pages={649--658},
  year={1992},
  publisher={Optical Society of America}
}

@article{ho2020denoising,
  title={Denoising diffusion probabilistic models},
  author={Ho, Jonathan and Jain, Ajay and Abbeel, Pieter},
  journal={Advances in neural information processing systems},
  volume={33},
  pages={6840--6851},
  year={2020}
}

@article{zhou2022learning,
  title={Learning stochastic object models from medical imaging measurements by use of advanced ambient generative adversarial networks},
  author={Zhou, Weimin and Bhadra, Sayantan and Brooks, Frank J and Li, Hua and Anastasio, Mark A},
  journal={Journal of Medical Imaging},
  volume={9},
  number={1},
  pages={015503--015503},
  year={2022},
  publisher={Society of Photo-Optical Instrumentation Engineers}
}

@inproceedings{xu2025ambient,
  title={Ambient-denoising diffusion generative-adversarial networks for establishing stochastic-object models from noisy-image data},
  author={Xu, Xichen and Chen, Wentao and Zhou, Weimin},
  booktitle={Medical Imaging 2025: Image Perception, Observer Performance, and Technology Assessment},
  volume={13409},
  pages={156--163},
  year={2025},
  organization={SPIE}
}

@article{daras2023ambient,
  title={Ambient diffusion: Learning clean distributions from corrupted data},
  author={Daras, Giannis and Shah, Kulin and Dagan, Yuval and Gollakota, Aravind and Dimakis, Alex and Klivans, Adam},
  journal={Advances in Neural Information Processing Systems},
  volume={36},
  pages={288--313},
  year={2023}
}

@article{yan2018deeplesion,
  title={DeepLesion: automated mining of large-scale lesion annotations and universal lesion detection with deep learning},
  author={Yan, Ke and Wang, Xiaosong and Lu, Le and Summers, Ronald M},
  journal={Journal of medical imaging},
  volume={5},
  number={3},
  pages={036501--036501},
  year={2018},
  publisher={Society of Photo-Optical Instrumentation Engineers}
}

@incollection{heath1998current,
  title={Current status of the digital database for screening mammography},
  author={Heath, Michael and Bowyer, Kevin and Kopans, Daniel and Kegelmeyer Jr, P and Moore, Richard and Chang, Kyong and Munishkumaran, S},
  booktitle={Digital mammography: nijmegen, 1998},
  pages={457--460},
  year={1998},
  publisher={Springer}
}

@article{lee2017curated,
  title={A curated mammography data set for use in computer-aided detection and diagnosis research},
  author={Lee, Rebecca Sawyer and Gimenez, Francisco and Hoogi, Assaf and Miyake, Kanae Kawai and Gorovoy, Mia and Rubin, Daniel L},
  journal={Scientific data},
  volume={4},
  number={1},
  pages={1--9},
  year={2017},
  publisher={Nature Publishing Group}
}

@article{kawar2022denoising,
  title={Denoising diffusion restoration models},
  author={Kawar, Bahjat and Elad, Michael and Ermon, Stefano and Song, Jiaming},
  journal={Advances in neural information processing systems},
  volume={35},
  pages={23593--23606},
  year={2022}
}

@article{chung2022diffusion,
  title={Diffusion posterior sampling for general noisy inverse problems},
  author={Chung, Hyungjin and Kim, Jeongsol and Mccann, Michael T and Klasky, Marc L and Ye, Jong Chul},
  journal={arXiv preprint arXiv:2209.14687},
  year={2022}
}

@article{li2024stimulating,
  title={Stimulating diffusion model for image denoising via adaptive embedding and ensembling},
  author={Li, Tong and Feng, Hansen and Wang, Lizhi and Zhu, Lin and Xiong, Zhiwei and Huang, Hua},
  journal={IEEE Transactions on Pattern Analysis and Machine Intelligence},
  year={2024},
  publisher={IEEE}
}

@article{gravel2004method,
  title={A method for modeling noise in medical images},
  author={Gravel, Pierre and Beaudoin, Gilles and De Guise, Jacques A},
  journal={IEEE Transactions on medical imaging},
  volume={23},
  number={10},
  pages={1221--1232},
  year={2004},
  publisher={IEEE}
}

@article{sharma2025detail,
  title={Detail-preserving denoising of CT and MRI images via adaptive clustering and non-local means algorithm},
  author={Sharma, Mohit and Dogra, Ayush and Goyal, Bhawna and Gupta, Anita and Saikia, Manob Jyoti},
  journal={Scientific Reports},
  volume={15},
  number={1},
  pages={23859},
  year={2025},
  publisher={Nature Publishing Group UK London}
}

@article{fartiyal2025dual,
  title={Dual Path Learning--learning from noise and context for medical image denoising},
  author={Fartiyal, Jitindra and Freire, Pedro and Whayeb, Yasmeen and Wolffsohn, James S and Turitsyn, Sergei K and Sokolov, Sergei G},
  journal={arXiv preprint arXiv:2507.19035
        
        
        
        
        
        },
  year={2025}
}

@inproceedings{xu2024ambientcyclegan,
  title={AmbientCycleGAN for establishing interpretable stochastic object models based on mathematical phantoms and medical imaging measurements},
  author={Xu, Xichen and Chen, Wentao and Zhou, Weimin},
  booktitle={Medical Imaging 2024: Image Perception, Observer Performance, and Technology Assessment},
  volume={12929},
  pages={234--240},
  year={2024},
  organization={SPIE}
}

@article{hung2023med,
  title={Med-cdiff: Conditional medical image generation with diffusion models},
  author={Hung, Alex Ling Yu and Zhao, Kai and Zheng, Haoxin and Yan, Ran and Raman, Steven S and Terzopoulos, Demetri and Sung, Kyunghyun},
  journal={Bioengineering},
  volume={10},
  number={11},
  pages={1258},
  year={2023},
  publisher={MDPI}
}

@article{cheng2021applications,
  title={Applications of artificial intelligence in nuclear medicine image generation},
  author={Cheng, Zhibiao and Wen, Junhai and Huang, Gang and Yan, Jianhua},
  journal={Quantitative Imaging in Medicine and Surgery},
  volume={11},
  number={6},
  pages={2792},
  year={2021}
}

@article{ambstylegan3,
  title={Learning stochastic object models from medical imaging measurements by use of advanced ambient generative adversarial networks},
  author={Zhou, Weimin and Bhadra, Sayantan and Brooks, Frank J and Li, Hua and Anastasio, Mark A},
  journal={Journal of Medical Imaging},
  volume={9},
  number={1},
  pages={015503--015503},
  year={2022},
  publisher={Society of Photo-Optical Instrumentation Engineers}
}

\end{document}